\documentclass[12pt,tightenlines,superscriptaddress, nofootinbib, amsmath, amssymb]{revtex4-1}
\usepackage{graphicx} 

\usepackage{caption}
\usepackage{subcaption}
\usepackage{ragged2e}  
\DeclareCaptionJustification{justified}{\justifying}

\usepackage{comment}

\newcommand\pubdate{\today}

\def\Title#1{\begin{center} {\Large #1 } \end{center}}
\def\Author#1{\begin{center}{ \sc #1} \end{center}}
\def\Address#1{\begin{center}{ \it #1} \end{center}}

\newcommand\pubblock{\rightline{\begin{tabular}{l}  \\ 
         \pubdate  \end{tabular}}}
\newenvironment{Abstract}{\begin{quotation}  }{\end{quotation}}
\newenvironment{Presented}{\begin{quotation} \begin{center} 
             PRESENTED AT\end{center}\bigskip 
      \begin{center}\begin{large}}{\end{large}\end{center} \end{quotation}}

\begin{document}
\begin{titlepage}
 \pubblock
\vfill
\Title{Single-Inclusive Particle Production from $pA$ Collision at Next-to-Leading Order}
\vfill
\Author{Heikki M\"antysaari}
\Author{Yossathorn Tawabutr}
\Address{Department of Physics, University of Jyv\"askyl\"a, P.O. Box 35, 40014 University of Jyv\"askyl\"a,
Finland}
\Address{Helsinki Institute of Physics, P.O. Box 64, 00014 University of Helsinki, Finland}
\vfill
\begin{Abstract}
We present the first fully consistent NLO calculation of the single-inclusive forward hadron production in proton-nucleus ($pA$) collisions under the color glass condensate (CGC) framework. In the dilute-dense limit, the NLO cross-section can be written as a convolution of the NLO impact factor, NLO parton distribution function (PDF), NLO fragmentation function (FF) and dipole-target scattering amplitude which satisfies the NLO small-$x$ Balitsky-Kovchegov (BK) evolution.

We demonstrate that, without the NLO corrections to the impact factor, we obtain a significant Cronin peak when the dipole amplitude satisfies the NLO BK equation. This would contradict the recent LHCb results \cite{LHCb:2022vfn}. However, the Cronin peak becomes suppressed when the NLO correction to the impact factor is included. This is the main result of this work. The dependence on resummation schemes for the NLO BK evolution will also be discussed.
\end{Abstract}
\vfill
\begin{Presented}
DIS2023: XXX International Workshop on Deep-Inelastic Scattering and
Related Subjects, \\
Michigan State University, USA, 27-31 March 2023 \\ 
     \includegraphics[width=9cm]{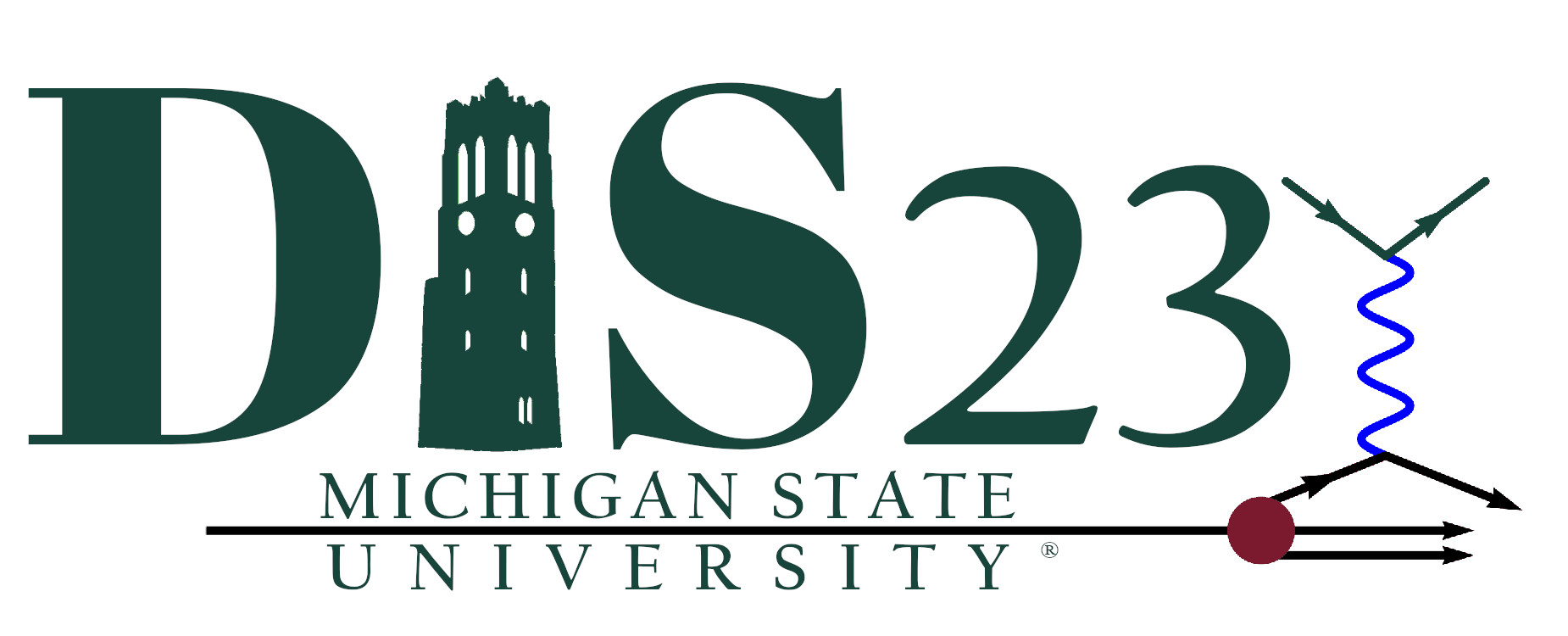}
\end{Presented}
\vfill
\end{titlepage}

\section{Introduction}

This article is based on the work presented in \cite{NLOsinc}, which is in preparation. 

Single-inclusive hadron productions in forward proton-proton ($pp$) or proton-nucleus ($pA$) collisions at high energy can be expressed using the CGC formalism \cite{Mueller:1989st,Mueller:1993rr,Balitsky:1995ub,Gelis:2010nm} in terms of the unintegrated gluon distribution \cite{Dumitru:2002qt,Dumitru:2005gt,Chirilli:2011km,Chirilli:2012jd}. This opens up an opportunity to compare CGC calculations against experimental measurements in order to probe the small-$x$ structure of protons and nuclei \cite{Stasto:2013cha,Watanabe:2015tja,Shi:2021hwx}. As a result, forward hadron productions in $pp$ and $pA$ collisions have been an active area of study for more than two decades \cite{Dumitru:2002qt,Dumitru:2005gt,Chirilli:2011km,Chirilli:2012jd,Stasto:2013cha,Watanabe:2015tja,Shi:2021hwx,Altinoluk:2011qy,Lappi:2013zma,Altinoluk:2014eka,Ducloue:2017dit,Liu:2019iml,Kang:2019ysm,Liu:2020mpy}.

Consider a collision in the center-of-mass frame between a proton and a nucleon from the nucleus, which could be another proton, such that a forward parton from the proton -- with a large longitudinal momentum fraction, $x_p$ -- interacts with a parton from the nucleus with small longitudinal momentum fraction, $X_g$. As the collision takes place, the forward parton receives a transverse momentum, $k_{\perp}$, but remains forward. Eventually, it fragments into a hadron. A direct calculation of the kinematics allows us to write
\begin{align}\label{xpxg}
&x_p = \frac{k_{\perp}}{\sqrt{s}}\,e^y\;\;\;\;\text{and}\;\;\;\;X_g = \frac{k_{\perp}}{\sqrt{s}}\,e^{-y}\,,
\end{align}
where $y$ is the rapidity and $s$ is the squared center-of-mass energy per nucleon of the $pA$ collision. In this ``dilute-dense'' framework with $k_{\perp}$ greater than the saturation momentum, $Q_s$, the ``hybrid formalism'' applies \cite{Lappi:2013zma,Kovchegov:2001sc}, allowing us to write the hadron production cross section as a convolution of PDFs -- $q_f(x_p)$ for quarks and $g(x_p)$ for gluons -- FFs -- $D_{h/f}(z)$ for quarks and $D_{h/g}(z)$ for gluons -- and the unintegrated gluon distribution, the Fourier transform of the dipole amplitude \cite{Mueller:1989st,Mueller:1993rr,Kovchegov:2012mbw}. At the leading order (LO), we have \cite{Dumitru:2002qt,Dumitru:2005gt,Chirilli:2011km,Chirilli:2012jd}
\begin{align}\label{LO}
\frac{\mathrm{d}\sigma_{pA\to hX}}{\mathrm{d}^2p_{\perp}\,dy} &= \int\frac{dz}{z^2} \int\frac{d^2x_0\,d^2x_1}{(2\pi)^2} \,e^{-i\underline{k}\cdot(\underline{x}_0-\underline{x}_1)}   \left[\sum_fx_pq_f(x_p)\,D_{h/f}(z)\,\frac{1}{N_c}\left\langle\text{tr}\left[V_{\underline{0}}V_{\underline{1}}^{\dagger}\right]\right\rangle(X_g) \right.\notag\\
&\;\;\;\;+ \left. x_pg(x_p)\,D_{h/g}(z)\,\frac{1}{N_c^2-1}\left\langle\text{Tr}\left[U_{\underline{0}}U_{\underline{1}}^{\dagger}\right]\right\rangle(X_g) \right] ,
\end{align}
where $N_c$ is the number of quark colors and $\left\langle\cdots\right\rangle(X_g)$ is the ``CGC averaging'' \cite{Mueller:1989st,Mueller:1993rr,Kovchegov:2012mbw} over the target nucleus's quantum state evaluated at $X_g$. Finally, with the notation that $\underline{x} = (x^1,x^2)$ is a transverse vector,
\begin{subequations}\label{WL}
\begin{align}
&V_{\underline{n}} \equiv V_{\underline{x}_n} = \mathcal{P}\,\exp\left[ig\int\limits_{-\infty}^{\infty}dx^-t^aA^{+a}(x^+=0,x^-,\underline{x}_n) \right] , \label{WLV} \\
&U_{\underline{n}} \equiv U_{\underline{x}_n} = \mathcal{P}\,\exp\left[ig\int\limits_{-\infty}^{\infty}dx^-T^aA^{+a}(x^+=0,x^-,\underline{x}_n) \right] , \label{WLU}
\end{align}
\end{subequations}
are the fundamental and adjoint light-cone Wilson lines, respectively, with $\mathcal{P}$ being the path-ordering operator. Throughout this article, we employ the light-cone coordinates such that $x^{\pm}=(x^0\pm x^3)/\sqrt{2}$. The first term in the square brackets of Eq. \eqref{LO} corresponds to the ``quark channel,'' while the second term corresponds to the ``gluon channel'' \cite{Dumitru:2002qt,Dumitru:2005gt,Chirilli:2011km,Chirilli:2012jd}. 

Eq. \eqref{LO} receives next-to-leading-order (NLO) corrections from an emission of a ``primary parton'' either before or after the interaction with the target. The resulting cross section follows from a direct calculation in the light-cone perturbation theory (LCPT) \cite{Lepage:1980fj,Brodsky:1989pv}. For single-inclusive cross-sections, we integrate over the transverse position of one of the two outgoing partons, while keeping track of the other parton as it fragments into a hadron \cite{Chirilli:2011km,Chirilli:2012jd}. This leads to 4 different NLO channels -- $qq$, $qg$, $gq$ and $gg$ -- denoting the incoming parton and the outgoing parton we track, respectively. The resulting NLO expression will be omitted in this article for brevity.\footnote{See \cite{Chirilli:2012jd} for the full expression and its derivation.}

The NLO correction introduced above only concerns the ``impact factor.'' Additionally, each of the PDF, FF and dipole amplitude that enter Eq. \eqref{LO} receive NLO corrections. For the dipole amplitude, the corrections come through the high-energy BK evolution \cite{Balitsky:1997mk,Kovchegov:1999yj,Kovchegov:1999ua,Balitsky:2007feb}. 

In this work, we perform for the first time the full-NLO calculation of single-inclusive hadron productions, including all NLO corrections outlined above \cite{NLOsinc}. The ingredients of our calculation are detailed below, followed by the preliminary results and the comparison with the recent LHCb's forward $p$Pb $\pi^0$ production data \cite{LHCb:2022vfn}.

\section{Ingredients}

As mentioned previously, the dipole amplitude for the $pp$ collision is taken from \cite{Beuf:2020dxl}, in which the NLO BK evolution \cite{Balitsky:2007feb,Beuf:2014uia,Iancu:2015vea,Ducloue:2019ezk} is applied to the initial condition given by the MV$^{\gamma}$ model,
\begin{align}\label{MVg}
&S^{(0)}(\underline{x}_0,\underline{x}_1) \equiv \frac{1}{N_c}\left\langle\text{tr}\left[V_{\underline{0}}V_{\underline{1}}^{\dagger}\right]\right\rangle(x_0) = \exp\left[ - \frac{1}{4}\left(x_{10}^2Q^2_{s,0}\right)^{\gamma}\ln\left(\frac{1}{x_{10}\Lambda} + e\right) \right]
\end{align}
at initial value, $x_0=0.01$. Here, $x_{10} = |\underline{x}_1-\underline{x}_0|$ and $\Lambda = 0.241$ GeV is the QCD scale, while $\gamma$ and $Q_{s,0}$ are the model parameters determined by fitting the evolved dipole amplitude to the HERA structure function data \cite{H1:2009pze,H1:2012xnw,H1:2015ubc,H1:2018flt}. Note that the addition by $e$ inside the logarithm is so that the infrared divergence is regulated. Since there are several schemes to resum the double-logarithmic terms in the NLO BK evolution, we follow the approach of \cite{Beuf:2020dxl} and perform the cross-section calculation separately for each resummation scheme, using the fitted parameters from \cite{Beuf:2020dxl} to obtain the dipole amplitude. A comparison of the resulting cross-section is given in Section \ref{sect:res}. 

For the $pA$ case, we employ the optical Glauber model introduced in \cite{Lappi:2013zma}, which gives the following initial condition for the dipole amplitude in the $pA$ case, 
\begin{align}\label{MVpA}
&S^{(0)}_{pA}(\underline{x}_0,\underline{x}_1;b_{\perp}) = \exp\left[ - \frac{1}{4}\,\frac{\sigma_0}{2}AT_A(b_{\perp})\left(x_{10}^2Q^2_{s,0}\right)^{\gamma}\ln\left(\frac{1}{x_{10}\Lambda} + e\right) \right] ,
\end{align}
where $\sigma_0/2$ is the transverse area of a proton and $A$ is the mass number of the nucleus. Here, $T_A(b_{\perp})$ is the transverse thickness function of the nucleus, which can be obtained from the Woods-Saxon distribution of nuclear density. Eq. \eqref{MVpA} depends on the impact parameter, $b_{\perp}$, of the $pA$ collision, in addition to the model parameters, $\gamma$ and $Q_{s,0}$. From there, the NLO BK evolution is applied separately for each $b_{\perp}$ to obtain the evolved dipole amplitude that are eventually used to calculate the particle production yield in the $pA$ collision as a function of $b_{\perp}$. Then, we integrate over $b_{\perp}$ weighted by the average number of binary collisions to obtain the overall $pA$ cross-section \cite{NLOsinc}. 

The dipole amplitude in each case is convoluted with the PDF and FF. For the PDF, we employ the Martin-Stirling-Thorne-Watt (MSTW) PDF at NLO \cite{Martin:2009iq} through the LHAPDF library \cite{Buckley:2014ana}. As for the FF, we use the de Florian-Sassot-Stratmann (DSS) results at NLO \cite{deFlorian:2007aj}.

Finally, the NLO impact factor has collinear and rapidity divergences. The former is subtracted by the DGLAP evolution of the PDF and FF \cite{Chirilli:2012jd}. In \cite{Chirilli:2012jd}, the rapidity divergences are subtracted by the LO BK evolution of the dipole amplitude. However, in \cite{Ducloue:2017dit}, it is shown that one could leave the rapidity divergence in the NLO impact factor while evaluating the LO impact factor terms at the initial condition, $X_g = x_0$. This is called the ``unsubtracted scheme,'' and it is theoretically more exact because it does not require subtracting and adding a potentially large contribution, which can cause problems when the running coupling is at play \cite{Ducloue:2017dit}. In this work, we employ the momentum-space running coupling prescription for the impact factor, making the unsubtracted scheme a better choice.

With all the ingredients specified above, we calculate the single-inclusive $\pi^0$ production cross-section in $p$Pb collisions at the full NLO level, whose results are presented in the next section. Note that this is a novel development. For the first time, the dipole amplitude fitted to the data using NLO BK evolution \cite{Beuf:2020dxl} is employed in such calculations.\footnote{In \cite{Stasto:2013cha,Watanabe:2015tja,Shi:2021hwx}, the NLO corrections to the BK evolution of the dipole amplitude are not included.}

\section{Results}\label{sect:res}

\subsection{Hadron Production Spectrum}

\begin{figure}
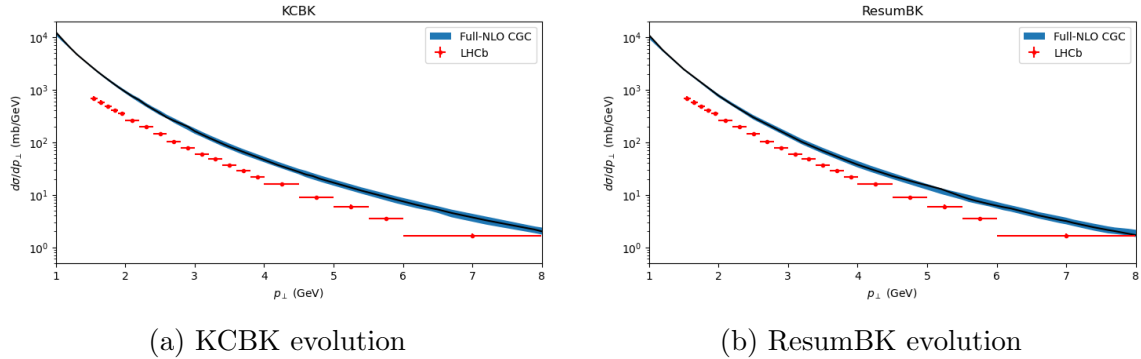

     \centering
     \begin{subfigure}[b]{0.48\textwidth}
         \centering
         \includegraphics[width=\textwidth]{spectra_KCBK.pdf}
         \caption{KCBK evolution}
         \label{fig:spectra_KCBK}
     \end{subfigure} 
     \;\;\;
     \begin{subfigure}[b]{0.48\textwidth}
         \centering
         \includegraphics[width=\textwidth]{spectra_ResumBK.pdf}
         \caption{ResumBK evolution}
         \label{fig:spectra_ResumBK}
     \end{subfigure}
     \caption{The plots of $\pi^0$ spectra at $\sqrt{s} = 8.16$ TeV $y=3$, versus the transverse momentum, $p_{\perp}$, of the outgoing $\pi^0$. In each plot, the blue band displays our full-NLO results using the specified resummation scheme for the NLO BK evolution, with factorization scale varied such that $\mu = 2p_{\perp},4p_{\perp},8p_{\perp}$. The red dots show the LHCb results, with the error bars combining both statistical and systematic uncertainties \cite{LHCb:2022vfn}.}
     \label{fig:spectra}
\end{figure}

We perform the calculation at LHCb's kinematics, with center-of-mass energy, $\sqrt{s} = 8.16$ TeV, and rapidity, $y=3$, using two different resummation schemes in the NLO BK evolution of the dipole: (i) kinematically-constrained BK (KCBK) \cite{Beuf:2014uia} and (ii) local-rapidity resummed BK (ResumBK) \cite{Iancu:2015vea}. Respectively, the resulting $\pi^0$ cross-sections for the two resummation schemes are shown in Figure \ref{fig:spectra}. There, the error bands are constructed by varying the factorization scale such that $\mu = 2p_{\perp},4p_{\perp},8p_{\perp}$.\footnote{In \cite{Stasto:2013cha}, the cross-section appears to be stable only for $\mu\gtrsim 2p_{\perp}$.} 

From Figure \ref{fig:spectra}, we see that our spectra differ very slightly across the resummation schemes. On a more unfortunate note, they significantly overestimate the LHCb results. However, the functional form seems to be similar, with the discrepancy coming mainly from an overall factor. We suspect that the mismatch may result from a problem when the model with parameters fitted from HERA data is generalized to $pA$ collisions using the optical Glauber model \cite{NLOsinc}. The issue will be studied in a future work. For the remainder of the article, we will only consider the $b_{\perp}=0$ case where the potential issues with $pA$ dipoles are not as severe.

\subsection{Nuclear Modification Factor: Cronin Effect}

Despite the mismatch in our $p$Pb spectra with the LHCb measurement, the most striking results of our calculation are in the nuclear modification factor, which is defined in the case of $b_{\perp}=0$ as
\begin{align}\label{RpA}
&R_{pA} = \frac{\mathrm{d}N_{pA\to hX}/\mathrm{d}^2p_{\perp}\mathrm{d}y}{\left[N_{\text{bin}}\big|_{b_{\perp}=0} \right] \mathrm{d}N_{pp\to hX}/\mathrm{d}^2p_{\perp}\mathrm{d}y} \, ,
\end{align}
where $N_{pp/pA\to hX}$ is the particle production yield and $N_{\text{bin}}\big|_{b_{\perp}=0}$ is the number of binary collisions in $pA$ collisions at $b_{\perp}=0$. The factor, $R_{pA}$, allows for a direct comparison between the $pp$ and $pA$ cross-sections, in such the way that $R_{pA}=1$ would imply that the nucleus behaved in the context of a $pA$ collision as if it were $A$ separate protons.

\begin{figure}
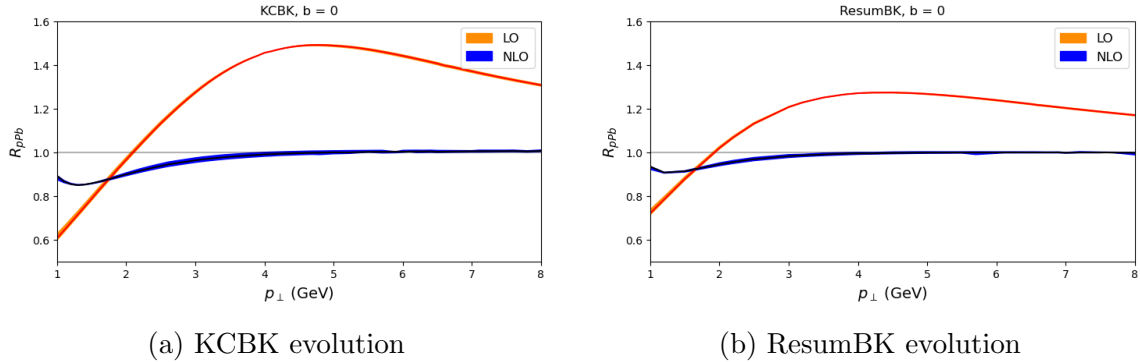

     \centering
     \begin{subfigure}[b]{0.48\textwidth}
         \centering
         \includegraphics[width=\textwidth]{RpA_KCBK.pdf}
         \caption{KCBK evolution}
         \label{fig:RpA_KCBK}
     \end{subfigure} 
     \;\;\;
     \begin{subfigure}[b]{0.48\textwidth}
         \centering
         \includegraphics[width=\textwidth]{RpA_ResumBK.pdf}
         \caption{ResumBK evolution}
         \label{fig:RpA_ResumBK}
     \end{subfigure}
     \caption{The plots of nuclear modification factor, $R_{p\text{Pb}}$, at $\sqrt{s} = 8.16$ TeV $y=3$, versus the transverse momentum, $p_{\perp}$, of the outgoing $\pi^0$. The calculation considers the $p$Pb collision at $b_{\perp}=0$ only. In each plot, the blue band is the full-NLO results, while the orange band is computed using LO impact factor, with the PDF, FF and BK evolution being at NLO for both cases. The bands themselves are constructed by varying the factorization scale such that $\mu = 2p_{\perp},4p_{\perp},8p_{\perp}$. }
     \label{fig:RpA}
\end{figure}

As mentioned above, we only consider the $pA$ collisions at $b_{\perp}=0$. The results for both resummation schemes are shown in Figure \ref{fig:RpA}. With LO impact factor but PDF, FF and dipole at NLO (the orange bands), we see a clear Cronin effect around $p_{\perp}\approx$ 4 - 5 GeV, which is larger with the KCBK evolution. However, in the full-NLO case (the blue bands), the Cronin peak disappears, and the discrepancy between KCBK and ResumBK results become much smaller. The former is especially desirable because the $R_{p\text{Pb}}$ measurement from LHCb displays no Cronin peak \cite{LHCb:2022vfn}. This result is of great importance -- if the NLO corrections to the dipole's evolution is to be included, then the NLO corrections must consistently be included everywhere else: the impact factor, PDF and FF.


\section{Conclusion and Outlook}

For the first time, we employ the CGC framework to compute the hadron production cross section in $pA$ collisions at the full NLO accuracy consistently with the DIS data. The main result of this work is that the NLO corrections to the impact factor are essential to remove the Cronin effect at moderate hadron’s transverse momentum, $p_{\perp}$. Furthermore, the discrepancy in $R_{pA}$ due to the NLO BK resummation scheme becomes suppressed in the full-NLO case where the impact factor is also at NLO.

There remains a significant discrepancy between our $pA$ spectra and the LHCb results \cite{LHCb:2022vfn}, possibly due to the dipole fit and its generalization to $pA$ collisions. The issue will be investigated further in a future work. In light of upcoming forward scattering measurements \cite{ALICE:2023fov}, the dependence of our results on the rapidity, $y$, will also be studied. Last but not least, as an additional cross-check, our calculation will be repeated with the target momentum fraction BK (TBK) evolution \cite{Ducloue:2019ezk}, which is another available resummation scheme of the NLO BK evolution.

\section{Acknowledgments}

YT would like to thank Dr. Tuomas Lappi for helpful discussions and the DIS2023 organizers for the opportunity to present the work.

The authors are supported by the Academy of Finland, the Centre of Excellence in Quark Matter, and projects 338263 and 346567, under the European Union’s Horizon 2020 research and innovation programme by the European Research Council (ERC, grant agreement No. ERC-2018-ADG-835105 YoctoLHC) and by the STRONG-2020 project (grant agreement No. 824093). The content of this article does not reflect the official opinion of the European Union and responsibility for the information and views expressed therein lies entirely with the authors.



\providecommand{\href}[2]{#2}\begingroup\raggedright\endgroup

\end{document}